\documentclass[prd,superscriptaddress,amsfonts,amssymb,amsmath,showpacs,twocolumn]{revtex4-2}
\usepackage{bm}
\usepackage{amsfonts}
\usepackage{latexsym}
\usepackage[latin1]{inputenc}
\usepackage{graphicx}
\usepackage{amsmath}
\usepackage{palatino}
\usepackage{mathpazo}
\usepackage{textcomp}
\usepackage{float}
\usepackage{booktabs}
\usepackage{dcolumn}
\usepackage{booktabs}
\usepackage{multirow}
\usepackage{hyperref}
\hypersetup{colorlinks,citecolor=blue}
\usepackage{amsmath}
\usepackage{xcolor}
\usepackage{orcidlink}
\usepackage[caption=false]{subfig}
\usepackage{commath}
\def\jnl@style{\it}
\def\aaref@jnl#1{{\jnl@style#1}}

\def\aaref@jnl#1{{\jnl@style#1}}

\def\aj{\aaref@jnl{AJ}}                   
\def\apj{\aaref@jnl{ApJ}}                 
\def\apjl{\aaref@jnl{ApJ}}                
\def\apjs{\aaref@jnl{ApJS}}               
\def\apss{\aaref@jnl{Ap\&SS}}             
\def\aap{\aaref@jnl{A\&A}}                
\def\aapr{\aaref@jnl{A\&A~Rev.}}          
\def\aaps{\aaref@jnl{A\&AS}}              
\def\mnras{\aaref@jnl{Mon.~Not.~Roy.~Astron.~Soc.}}             
\def\prd{\aaref@jnl{Phys.~Rev.~D}}        
\def\prc{\aaref@jnl{Phys.~Rev.~C}}  
\def\prl{\aaref@jnl{Phys.~Rev.~Lett.}}    
\def\qjras{\aaref@jnl{QJRAS}}             
\def\skytel{\aaref@jnl{S\&T}}             
\def\ssr{\aaref@jnl{Space~Sci.~Rev.}}     
\def\zap{\aaref@jnl{ZAp}}                 
\def\nat{\aaref@jnl{Nature}}              
\def\aplett{\aaref@jnl{Astrophys.~Lett.}} 
\def\apspr{\aaref@jnl{Astrophys.~Space~Phys.~Res.}} 
\def\physrep{\aaref@jnl{Phys.~Rep.}}      
\def\physscr{\aaref@jnl{Phys.~Scr}}       
\def\commat{\aaref@jnl{Comm.~Math.~Phys.}}              
\def\science{\aaref@jnl{Science}}               
\def\cqg{\aaref@jnl{Classical Quant.~Grav.}}            
\def\jpcs{\aaref@jnl{JPCS}}                                     
\def\ijmpd{\aaref@jnl{Int.~J.~Mod.~Phys.~D}}                    
\def\grg{\aaref@jnl{Gen.~Relat.~Gravit.}}               
\def\rpp{\aaref@jnl{Rep.~Prog.~Phys.}}          
\def\npa{\aaref@jnl{Nucl.~Phys.~A}}        
\def\lrr{\aaref@jnl{Living Rev.~Rel.}}                   
\def\jcap{\aaref@jnl{J.~Cosmology Astropart.~Phys.}}    
\def\rmp{\aaref@jnl{Rev.~Mod.~Phys.}}   
\def\epjc{\aaref@jnl{Eur.~Phys.~J.~C}}

\allowdisplaybreaks[1]

\begin{document}

\color{black}       

\title{Quintessence-like features in the late-time cosmological evolution of $f(Q)$ symmetric teleparallel gravity}

\author{N. Myrzakulov\orcidlink{0000-0001-8691-9939}}
\email[Email: ]{nmyrzakulov@gmail.com}
\affiliation{L. N. Gumilyov Eurasian National University, Astana 010008,
Kazakhstan.}
\affiliation{Ratbay Myrzakulov Eurasian International Centre for Theoretical
Physics, Astana 010009, Kazakhstan.}

\author{M. Koussour\orcidlink{0000-0002-4188-0572}}
\email[Email: ]{pr.mouhssine@gmail.com}
\affiliation{Quantum Physics and Magnetism Team, LPMC, Faculty of Science Ben
M'sik,\\
Casablanca Hassan II University,
Morocco.}

\author{A. Mussatayeva \orcidlink{0000-0000-0000-0000}}
\email[Email: ]{a.b.mussatayeva@gmail.com}
\affiliation{
Department of Physics and Chemistry, S. Seifullin Kazakh Agrotechnical University, Astana 010011, Kazakhstan}

\date{\today}

\begin{abstract}
In this study, we investigate the cosmological history within the framework of modified $f(Q)$ gravity, which proposes an alternative theory of gravity where the gravitational force is described by a non-metricity scalar. By employing a parametrization scheme for the Hubble parameter, we obtain the exact solution to the field equations in $f(Q)$ cosmology. To constrain our model, we utilize external datasets, including 57 data points from the Hubble dataset, 1048 data points from the SN dataset, and six data points from the BAO dataset. This enables us to determine the best-fit values for the model parameters involved in the parameterization scheme. We analyze the cosmic evolution of various cosmological parameters, including the deceleration parameter, which exhibits the expected behavior in late-time cosmology and changes its signature with redshift. Further, we present the evolution of energy density, EoS parameter, and other geometrical parameters with respect to redshift. Furthermore, we discuss several cosmological tests and diagnostic analyses. Our findings demonstrate that the late-time cosmic evolution can be adequately described without the need for dark energy by employing a parametrization scheme in modified gravity.
\end{abstract}

\maketitle

\section{Introduction}

\label{sec1}

Our universe is expanding at an accelerated rate, as supported by cosmological evidence from various sources such as Type Ia Supernovae (SN), Baryonic Acoustic Oscillations (BAO), and the Cosmic Microwave Background (CMB) \cite%
{Riess,Perlmutter,C.L.,D.N., D.J.,W.J.}. One of the most significant challenges in modern physics is comprehending the underlying cause of cosmic acceleration. Within the framework of General Relativity (GR), it is hypothesized that this acceleration is attributed to an enigmatic form of energy known as Dark Energy (DE). The prevailing belief is that DE dominates the Universe; however, none of the existing hypotheses regarding DE provide a complete explanation for this phenomenon. The most successful cosmological model accounting for DE is the $\Lambda$CDM model, commonly referred to as the cosmological constant ($\Lambda$) model, where the equation of state (EoS) parameter $\omega_{\Lambda}=-1$. 
The physical nature of the cosmological constant is linked to the existence of non-zero vacuum energy and can be estimated through quantum field theory. However, the theoretical predictions and observational cosmological data differ significantly by several orders of magnitude, presenting a profound fine-tuning challenge \cite{S.W.}. 

In an attempt to address the cosmological constant problem, various ideas involving DE with a time-varying energy density have been proposed in the literature \cite{A5, A6, A7}. Modified gravity is another theory aimed at explaining the present acceleration of the universe. One straightforward approach is $f(R)$ gravity \cite{A8, A9,My1}, where the scalar curvature $R$ in Einstein's Lagrangian density is replaced by an arbitrary function of $R$, modifying the gravitational action of GR to include a generic scalar curvature function. Other theories, such as $f(R,T)$ \cite{A10, A11} and $f(R,L_{m})$ \cite{A12, A13}, where $T$ and $L_{m}$ represent the trace of the stress-energy tensor and matter Lagrangian, respectively, are based on a non-minimal interaction between matter and geometry. In GR, the Levi-Civita connection is employed as an affine connection to describe the gravitational interaction in Riemannian spacetime. However, there exist numerous possibilities for affine connections on any manifold, leading to various equivalent descriptions of gravity \cite{A14,My2}. In order to incorporate torsion $T$ and provide an alternative theory for gravitational interaction, the Teleparallel Equivalent of GR (TEGR) was proposed \cite{A15}. In TEGR, the Weitzenbock connection is utilized, which results in zero curvature $R$ and non-metricity $Q$. To further reduce the constraint on non-metricity while still eliminating both curvature and torsion, a novel theory called the Symmetric Teleparallel Equivalent of GR (STEGR) has been developed \cite{A16}. Within this theory, one can formulate $f(Q)$ gravity, where the non-metricity $Q$ describes the gravitational interaction \cite{A17, Jim}. Recently, researchers have shown increased interest in modified theories of $f(Q)$ gravity as they offer insights into the phenomena observed in the Universe. Consequently, an extension of $f(Q)$ gravity known as $f(Q,T)$ gravity has been proposed \cite{A18}. Similar to $f(R,T)$ gravity, $f(Q,T)$ gravity is based on a non-minimal coupling between the non-metricity $Q$ and the trace of the energy-momentum tensor $T$, rather than the Ricci scalar $R$. Several investigations have provided alternative explanations for the current cosmic acceleration of the universe and have offered viable solutions to the DE problem through $f(Q)$ gravity \cite{A21, MK2, MK3}. A relevant perspective in line with this paper is presented in \cite{A23}, where the parametrization technique is employed to explore cosmological models within the framework of $f(Q)$ gravity. Furthermore, references \cite{A24, A25} presented the initial cosmological solutions in $f(Q)$ gravity, while \cite{A27} investigated the corresponding energy conditions in $f(Q)$ gravity. The parametrization method not only allows us to investigate $f(Q)$ cosmology by parametrizing the Hubble parameter but also offers a powerful framework for studying the underlying dynamics of the universe and the nature of DE. By parametrizing the Hubble parameter, we can express it as a function of a set of parameters that capture the essential features of the cosmological model under consideration. This parametric form provides a flexible description of the expansion rate of the universe, allowing us to explore a wide range of possibilities beyond the standard cosmological models. The referenced studies \cite{A28, A29, A30} highlight the significance of the parametrization technique in cosmology. These investigations have utilized various parameterizations to explore different aspects of the universe, ranging from the nature of DE to the modified gravity theories. The key advantage of this method is its ability to incorporate observational data in examining cosmological models \cite{A23}.

In this study, we consider the Hubble parameterization proposed in \cite{A31} and analyze the FLRW Universe within the framework of $f(Q)$ gravity theory using the functional form $f(Q) = -Q + \frac{\alpha}{Q}$, where $\alpha$ is a free model parameter. The article is organized as follows: Sec. \ref{sec2} introduces the formalism of $f(Q)$ gravity. Sec. \ref{sec3} presents the motion equations in the FLRW Universe. In Sec. \ref{sec4}, we analyze the cosmological model and derive expressions for the cosmological parameters. Sec. \ref{sec5} discusses the observational data and the methodology used to constrain the model parameters. The behavior of cosmological parameters, such as the deceleration parameter, density, and EoS parameter, is examined in Sec. \ref{sec6}. In addition, in Sec. \ref{sec7}, we investigate the energy conditions of the obtained solutions. In Sec. \ref{sec8}, we analyze the behavior of diagnostic parameters. Finally, our findings are briefly discussed in Sec. \ref{sec9}, which provides the concluding remarks.

\section{$f(Q)$ gravity theory}
\label{sec2}

In this section, we will construct the field equations of $f(Q)$ gravity, which requires an understanding of several key geometric concepts necessary for formulating a relativistic theory of gravitation. To initiate the discussion, we define the action of $f(Q)$ gravity as presented in \cite{A17},
\begin{equation}
S=\int {\frac{1}{2}f(Q)\sqrt{-g}d^{4}x}+\int {L_{m}\sqrt{-g}d^{4}x.}
\label{eq1}
\end{equation}

In this scenario, ${L_{m}}$ represents the matter Lagrangian, and $g$ denotes the determinant of the metric tensor $g_{\mu \nu}$. Moreover, $f(Q)$ is an arbitrary function of the non-metricity $Q$. The non-metricity tensor $Q_{\gamma \mu \nu}$, which is the covariant derivative of the metric tensor with respect to the Weyl-Cartan connection, is discussed in \cite{A32},
\begin{equation}
Q_{\gamma \mu \nu }=-\nabla _{\gamma }g_{\mu \nu }=-\partial _{\gamma
}g_{\mu \nu }+g_{\nu \sigma }\Upsilon {^{\sigma }}_{\mu \gamma }+g_{\sigma
\mu }\Upsilon {^{\sigma }}_{\nu \gamma }.
\end{equation}

The Christoffel symbol ${\Gamma ^{\gamma }}_{\mu \nu }$, the contortion tensor ${C^{\gamma }}_{\mu \nu }$, and the disformation tensor ${L^{\gamma }}_{\mu \nu }$ can be combined to form the Weyl-Cartan connection. This connection is defined in the following manner:
\begin{equation}
{\Upsilon ^{\gamma }}_{\mu \nu }={\Gamma ^{\gamma }}_{\mu \nu }+{C^{\gamma }}%
_{\mu \nu }+{L^{\gamma }}_{\mu \nu }.  \label{eq2}
\end{equation}

The first term in the previous expression, the Christoffel symbol, corresponds to the Levi-Civita connection associated with the metric $g_{\mu \nu}$. It is given by the following expression:
\begin{equation}
{\Gamma ^{\gamma }}_{\mu \nu }\equiv \frac{1}{2}g^{\gamma \sigma }\left(
\partial _{\mu }g_{\sigma \nu }+\partial _{\nu }g_{\sigma \mu }-\partial
_{\sigma }g_{\mu \nu }\right) .
\end{equation}

The torsion tensor ${T^{\gamma }}_{\mu \nu}$ is employed in the construction of the contortion tensor ${C^{\gamma }}_{\mu \nu}$, which is defined as
\begin{equation}
{T^{\gamma }}_{\mu \nu }\equiv \frac{1}{2}\left( \Upsilon {^{\gamma }}_{\mu
\nu }-\Upsilon {^{\gamma }}_{\nu \mu }\right) .
\end{equation}

According to the given definition, the contortion tensor is defined as
\begin{equation}
{C^{\gamma }}_{\mu \nu }\equiv \frac{1}{2}{(T^{\gamma }}_{\mu \nu}+T_{\mu \
\nu }^{\ \gamma }+T_{\nu \ \mu }^{\ \gamma })=-{C^{\gamma }}_{\nu \mu }.
\end{equation}

The disformation tensor is,%
\begin{eqnarray}
{L^{\gamma }}_{\mu \nu } &=&-\frac{1}{2}g^{\gamma \sigma }\left( \nabla
_{\nu }g_{\mu \sigma }+\nabla _{\mu }g_{\nu \sigma }-\nabla _{\gamma }g_{\mu
\nu }\right) , \\
&=&\frac{1}{2}g^{\gamma \sigma }\left( Q_{\nu \mu \sigma }+Q_{\mu \nu \sigma
}-Q_{\gamma \mu \nu }\right) ,  \notag \\
&=&{L^{\gamma }}_{\nu \mu }.
\end{eqnarray}

Based on this, the non-metricity can be expressed as
\begin{equation}
Q=-g^{\mu \nu }\left( {L^{\alpha }}_{\beta \mu }{L^{\beta }}_{\nu \alpha }-{%
L^{\alpha }}_{\beta \alpha }{L^{\beta }}_{\mu \nu }\right) .
\end{equation}

In the coincident gauge, when the covariant derivative is reduced to the partial derivative ($\nabla
_{\mu }\rightarrow \partial _{\mu }$), the non-metricity invariant is chosen to be equal to the Einstein Lagrangian. This gauge choice is consistent with symmetric teleparallel gravity and is known as the coincident gauge \cite{A33}. In the connection given by Eq. (\ref{eq2}), the non-metricity tensor, Levi-Civita connection, and torsion tensor are utilized to describe the disformation, curvature, and contortion, respectively. In the case of $f(Q)$ gravity, where torsion and curvature vanish, the connection reduces to solely the disformation within the coincident gauge.

The trace of the non-metricity tensor is given by
\begin{equation}
Q_{\alpha }=Q_{\alpha }{}^{\mu }{}_{\mu }\text{ \ \ and \ \ }\tilde{Q}%
_{\alpha }=Q^{\mu }{}_{\alpha \mu }.  \label{2k}
\end{equation}

The superpotential tensor, which is the conjugate of the non-metricity, is given by
\begin{equation}
4P^{\gamma }{}_{\mu \nu }=-Q^{\gamma }{}_{\mu \nu }+2Q_{(\mu }{}^{\gamma
}{}_{\nu )}+(Q^{\gamma }-\tilde{Q}^{\gamma })g_{\mu \nu }-\delta _{(\mu
}^{\gamma }Q_{\nu )}.  \label{2l}
\end{equation}

The non-metricity scalar can be obtained through the following expression:
\begin{equation}
Q=-Q_{\gamma \mu \nu }P^{\gamma \mu \nu }.  \label{2m}
\end{equation}

The content of the Universe is described as a perfect fluid matter with the following energy-momentum tensor,
\begin{equation}
T_{\mu \nu }=\frac{-2}{\sqrt{-g}}\frac{\delta (\sqrt{-g}L_{m})}{\delta
g^{\mu \nu }}.  \label{2n}
\end{equation}

By varying the action (\ref{eq1}) with respect to the components of the metric tensor $g_{\mu \nu}$, we obtain the following equations,
\begin{widetext}
\begin{equation}\label{2p}
\frac{2}{\sqrt{-g}}\nabla_\gamma (\sqrt{-g}f_Q P^\gamma\:_{\mu\nu}) + \frac{1}{2}g_{\mu\nu}f+f_Q(P_{\mu\gamma\beta}Q_\nu\:^{\gamma\beta} - 2Q_{\gamma\beta\mu}P^{\gamma\beta}\:_\nu) = -T_{\mu\nu}.
\end{equation}
\end{widetext}

In this case, we will consider $f_{Q}=\frac{df}{dQ}$, which represents the derivative of $f$ with respect to $Q$.

\section{Motion equations in $f\left( Q\right) $\ gravity}

\label{sec3}

In this context, we assume that the Universe can be described by the Friedmann-Lema\^{i}tre-Robertson-Walker (FLRW) metric, which is characterized by its homogeneity, isotropy, and spatial flatness.
\begin{equation}
ds^{2}=-dt^{2}+a^{2}(t)[dx^{2}+dy^{2}+dz^{2}],  \label{eq3a}
\end{equation}%

where $a(t)$ denotes the scale factor of the Universe. Moreover, the non-metricity scalar associated with the metric (\ref{eq3a}) is given by
\begin{equation}
Q=6H^{2},  \label{3b}
\end{equation}%
where $H$ represents the Hubble parameter, which quantifies the rate of expansion of the Universe.

In cosmology, the commonly used energy-momentum tensor is that of a perfect cosmic fluid, neglecting viscosity effects. In this case, the energy-momentum tensor is given by
\begin{equation}
T_{\mu \nu }=(\rho +p)u_{\mu }u_{\nu }+pg_{\mu \nu },  \label{3c}
\end{equation}%
where $\rho$ and $p$ represent the energy density and isotropic pressure of the perfect cosmic fluid, respectively, and $u^{\mu}=(1,0,0,0)$ represents the components of the four-velocity vector characterizing the fluid.

The modified Friedmann equations that govern the dynamics of the Universe in $f(Q)$ gravity are expressed as \cite{A33, A34}
\begin{equation}
3H^{2}=\frac{1}{2f_{Q}}\left( -\rho +\frac{f}{2}\right) ,  \label{F1}
\end{equation}%
and%
\begin{equation}
\dot{H}+3H^{2}+\frac{\dot{f}_{Q}}{f_{Q}}H=\frac{1}{2f_{Q}}\left( p+\frac{f}{2%
}\right) ,  \label{F2}
\end{equation}%
where an overdot denotes the derivative with respect to cosmic time $t$. It should be noted that the standard Friedmann equations of GR are recovered when assuming the function $f(Q)=-Q$ \cite{A34}.

Now, let us derive the matter/energy conservation equation in its well-known form,
\begin{equation}
\dot{\rho}+3H\left( \rho +p\right) =0  \label{3f}
\end{equation}

Using Eqs. (\ref{F1}) and (\ref{F2}), we can express the cosmic energy density $\rho$ and isotropic pressure $p$ of the fluid as
\begin{equation}
\rho =\frac{f}{2}-6H^{2}f_{Q},  \label{F22}
\end{equation}%
\begin{equation}
p=\left( \dot{H}+3H^{2}+\frac{\dot{f_{Q}}}{f_{Q}}H\right) 2f_{Q}-\frac{f}{2}.
\label{F33}
\end{equation}

\section{Cosmological $f(Q)$ Model}
\label{sec4}

In our current analysis, we consider the following functional form,

\begin{equation*}
f\left( Q\right) =-Q+\frac{\alpha }{Q},
\end{equation*}%
where $\alpha$ is the free model parameter. Consequently, we find $f_Q = -1 - \frac{\alpha}{Q^2}$. It is worth noting that for $\alpha = 0$, the model reduces to the well-established case of GR found in the literature \cite{A34}. The proposed functional form of $f(Q)$ has been extensively studied \cite{A22}, and it provides a modification to late-time cosmology, potentially giving rise to DE \cite{Jim}. Then, for this specific model of $f(Q)$, the equations for the energy density of the Universe, $\rho$, and the isotropic pressure, $p$, are as follows:
\begin{equation}
\rho =3H^{2}+\frac{\alpha }{4H^{2}},
\end{equation}%
and 
\begin{equation}
p=-3H^{2}-2\overset{.}{H}+\frac{\alpha }{2H^{2}}\left( \frac{\overset{.}{H}}{%
3H^{2}}-\frac{\alpha }{2}\right) .
\end{equation}

Furthermore, the EoS parameter can be obtained by taking
$\omega =\frac{p}{\rho }$,
\begin{equation}
\omega =-1+\frac{2\overset{.}{H}}{3H^{2}}-\frac{16\overset{.}{H}H^{2}}{%
12H^{4}+\alpha }.  \label{EoS}
\end{equation}

As observed, Eqs. (\ref{F22}) and (\ref{F33}) form a system of equations with $\rho$, $p$, and $H$ as the three unknowns. To obtain the exact solutions, we require an additional physically reasonable condition. In this study, we consider the parametrization of the Hubble parameter in terms of redshift as investigated in \cite{A31},
\begin{equation}
H\left( z\right) =H_{0}\left( \frac{m+\left( 1+z\right) ^{n}}{m+1}\right) ^{%
\frac{3}{2n}},  \label{Hz}
\end{equation}%
where $m$ and $n$ are free model parameters obtained from observational constraints, and $H_0$ represents the current value of the Hubble parameter.  The chosen parametrization for $H(z)$ is valid from the matter epoch ($z \gg 1$) to the infinite future ($z = -1$). Moreover, we assume that $q(z \gg 1) = \frac{1}{2}$ based on the appropriate form of the deceleration parameter (see Eq. (\ref{qz})), which is required for the formation of cosmic structures, and $q(z = -1) = -1$ for thermodynamic reasons. This proposed parametrization for $H(z)$ is consistent with a spatially flat $\Lambda$CDM model \cite{A31}. In this study, we aim to constrain the model parameters $H_0$, $m$, and $n$ using the most recent cosmological dataset. Although the model parameter $\alpha$ is not explicitly present in the expression for the Hubble parameter, we choose a specific value, $\alpha = 2$, in order to investigate the evolution of density, pressure, and EoS parameter and make predictions about late-time acceleration.

\section{Observational constraints}
\label{sec5}

In this section, we aim to constrain the model parameters ($H_0$, $m$, and $n$) of our obtained model using available external datasets. We utilize the Hubble $H(z)$ dataset, which consists of 57 data points, the Pantheon sample of SN dataset with 1048 data points, and the BAO dataset with six data points. To perform the parameter estimation, we employ the emcee Python library \cite{Mackey/2013}, which utilizes Bayesian analysis and likelihood functions along with the Markov Chain Monte Carlo (MCMC) approach. 

\subsection{$H(z)$ dataset}

The Hubble parameter can be expressed as $H(z)=-\frac{dz}{dt(1+z)}$. The model-independent value of the Hubble parameter can be determined by measuring the quantity $dt$, while $dz$ is obtained from a spectroscopic survey. We consider a dataset consisting of 57 data points within the redshift range $0.07 \leq z \leq 2.41$, which were collected using the differential age technique and line of sight BAO \cite{Sharov/2018}. To estimate the model parameters $m$ and $n$, we employ the chi-square function to calculate the mean values based on the dataset,
\begin{equation}
\chi _{H}^{2}(H_{0},m,n)=\sum\limits_{k=1}^{57}\frac{%
[H_{th}(H_{0},m,n,z_{k})-H_{obs}(z_{k})]^{2}}{\sigma _{H(z_{k})}^{2}},  \label{4a}
\end{equation}%
where, $H_{th}$ represents the predicted value of the Hubble parameter from the model, while $H_{obs}$ represents its observed value. The standard error in the measured data of $H(z)$ is denoted by $\sigma_{H(z_{k})}$.

Fig. \ref{h-z} depicts a comprehensive comparison between our proposed model and the well-established $\Lambda$CDM model in the field of cosmology. For the plot, we have considered specific cosmological parameters, namely $\Omega _{m_{0}}=0.315$ and $H_{0}=67.4$ $km/s/Mpc$ \cite{Planck}. These parameters serve as the foundational assumptions for the $\Lambda$CDM model. To provide a holistic view, the figure incorporates the experimental results from the Hubble dataset, which consists of an impressive collection of 57 data points. By including the Hubble results, Fig. \ref{h-z} facilitates a direct and illuminating comparison between our proposed model and the widely-accepted $\Lambda$CDM model. This allows researchers and cosmologists to assess the performance and validity of both models in light of the empirical data. The figure serves as a visual tool for analyzing the discrepancies, agreements, and overall consistency between the two models. It enables the identification of regions where the models align closely with the observational data and highlights areas where there are deviations or variations.

\begin{widetext}

\begin{figure}[h]
\centerline{\includegraphics[scale=0.55]{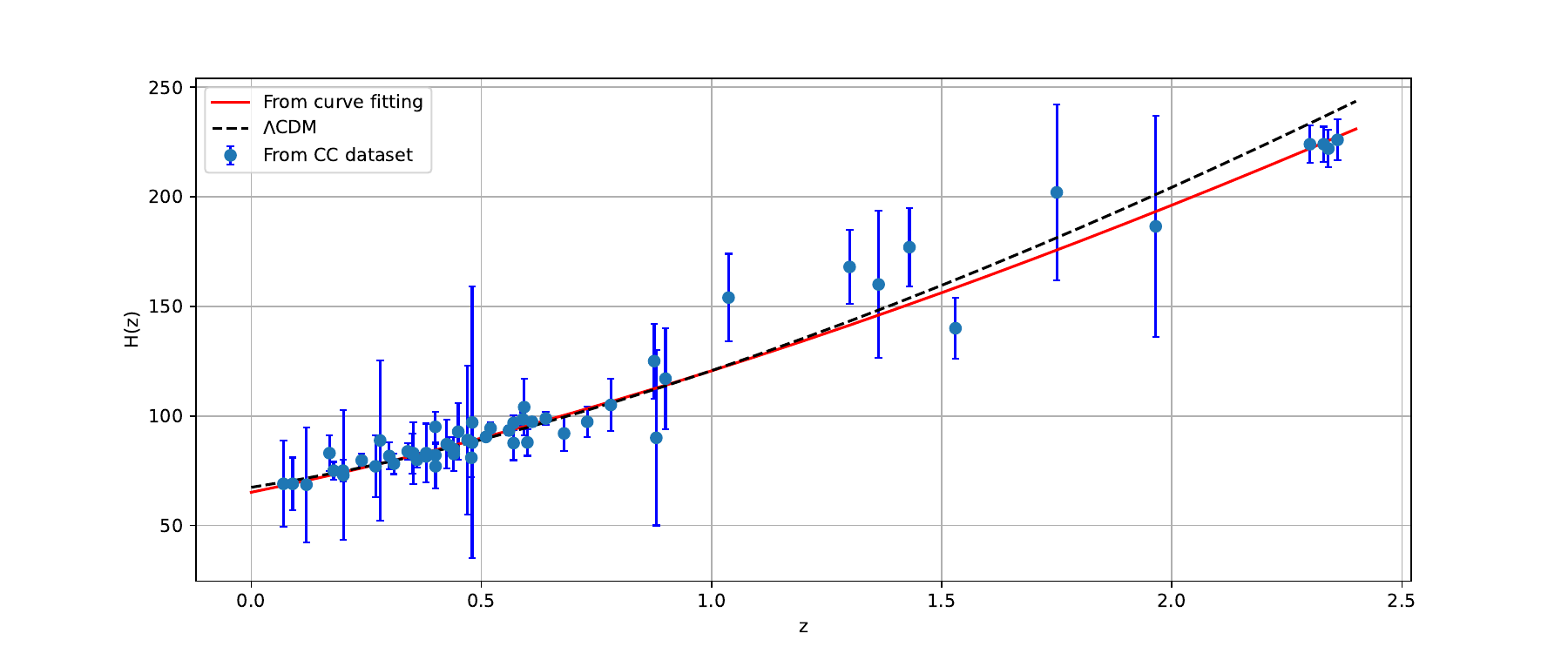}}
\caption{Error bar graph of $H$ versus $z$ for the assumed model. The curve for the model is shown by the solid red line, while the $\Lambda$CDM model is shown by the black dotted line. The 57 Hubble data points are represented by blue dots.}
\label{h-z}
\end{figure}

\begin{figure}[h]
\centerline{\includegraphics[scale=0.55]{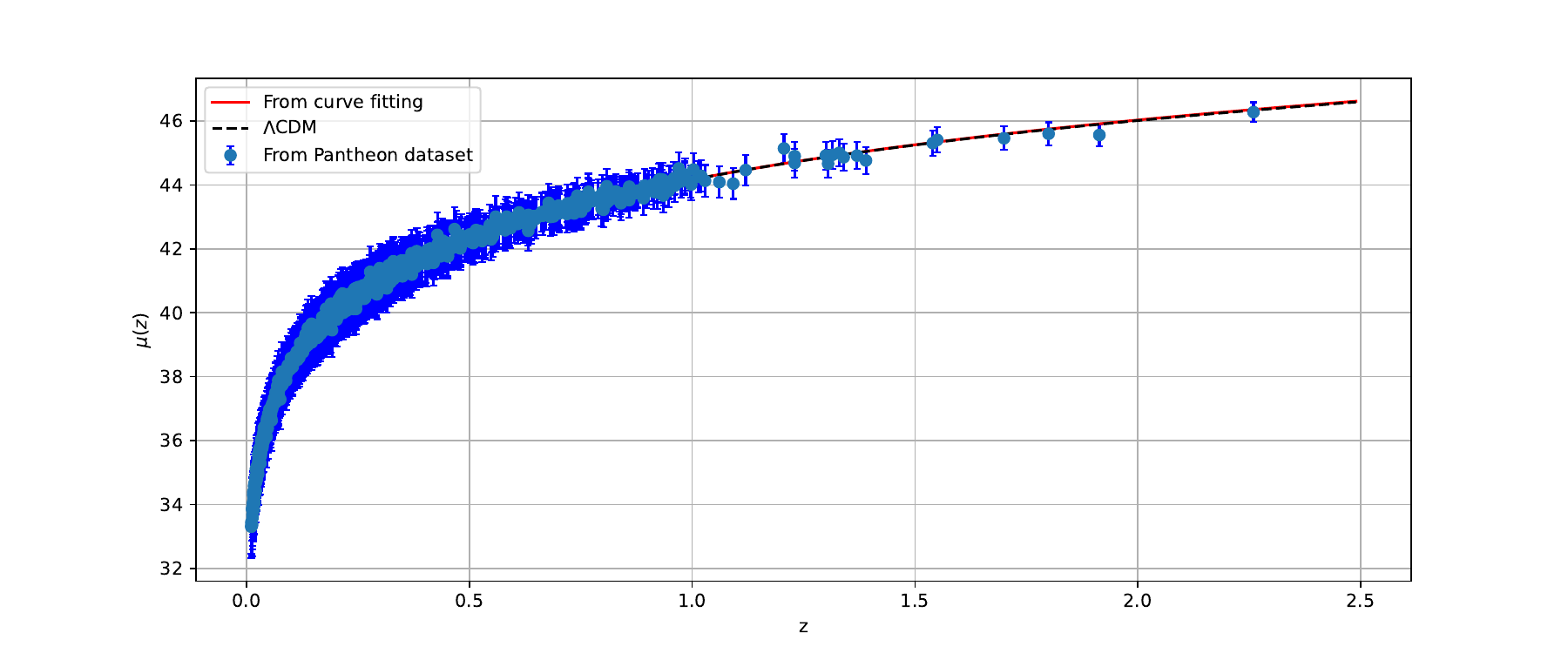}}
\caption{Error bar graph of $\mu(z)$ versus $z$ for the assumed model. The curve for the model is shown by the solid red line, while the $\Lambda$CDM model is shown by the black dotted line. The 1048 points of the Pantheon are represented by the blue dots with corresponding error bars.}
\label{mu-z}
\end{figure}

\end{widetext}

\subsection{SN dataset}

$1048$ data points from Pantheon supernovae type Ia samples data have
recently been made available. It includes data from the PanSTARSS1 Medium,
SDSS, SNLS, Deep Survey, a number of low redshift studies, and HST surveys.
The Pantheon supernovae type Ia samples, 1048 of which are in the redshift
range $0.01<z<2.3$, were assembled by Scolnic et al. \cite{Scolnic/2018}.
The luminosity distance for a spatially flat Universe is given by the
expression%
\begin{equation}
D_{L}(z)=(1+z)\int_{0}^{z}\frac{cdz^{\prime }}{H(z^{\prime })},  \label{4f}
\end{equation}%
where $c$ represents the speed of light.

The $\chi ^{2}$ function for type Ia supernovae is generated by correlating
the theoretical distance modulus%
\begin{equation}
\mu (z)=5 log_{10}D_{L}(z)+\mu _{0},  \label{4g}
\end{equation}%
with%
\begin{equation}
\mu _{0}=5 log(1/H_{0}Mpc)+25,  \label{4h}
\end{equation}%
like that%
\begin{equation}
\chi _{SN}^{2}(H_{0},m,n)=\sum_{k=1}^{1048}\dfrac{\left[ \mu _{obs}(z_{k})-\mu
_{th}(H_{0},m,n,z_{k})\right] ^{2}}{\sigma ^{2}(z_{k})}.  \label{4i}
\end{equation}%
Here, $\mu _{th}$ is the theoretical value of the distance modulus, $\mu
_{obs}$ is the observed value, and $\sigma ^{2}(z_{k})$ is the standard error of the observed value.

By including the Pantheon experimental results, Fig. Fig. \ref{mu-z} facilitates a direct and illuminating comparison between our proposed model and the widely-accepted $\Lambda$CDM model. 

\subsection{BAO dataset}

The BAO (Baryon Acoustic Oscillations) is well described in observational
cosmology. The fluctuations in the density of ordinary (baryonic) matter in the Universe are measured, which is influenced by acoustic density waves in
the early Universe's primordial plasma. Here, we have considered the six BAO
measurements at different redshifts for the 6dFGS, SDSS, and WiggleZ survey
as shown in (Tab. \ref{tab1}). The sound horizon, $r_{s}$, at the photon
decoupling epoch, $z_{\ast }$, determines the characteristic scale of BAO
and is provided by the relation:%
\begin{equation}
r_{s}(z_{\ast })=\frac{c}{\sqrt{3}}\int_{0}^{\frac{1}{1+z_{\ast }}}\frac{da}{%
a^{2}H(a)\sqrt{1+(3\Omega _{b0}/4\Omega _{\gamma 0})a}},  \label{4b}
\end{equation}%
where the current densities of baryons and photons are indicated here by the
symbols $\Omega _{b0}$ and $\Omega _{\gamma 0}$, respectively. In order to
measure BAO, the following relationships are used,%
\begin{equation}
\triangle \theta =\frac{r_{s}}{d_{A}(z)},  \label{4c}
\end{equation}%
\begin{equation}
d_{A}(z)=\int_{0}^{z}\frac{dz^{\prime }}{H(z^{\prime })},
\end{equation}%
\begin{equation}
\triangle z=H(z)r_{s}.
\end{equation}%
Here $\triangle \theta $ represents the observed angular separation, $d_{A}$
represents the measured angular diameter distance, and $\triangle z$
represents the measured redshift separation of the BAO characteristic in the
two-point correlation function of the galaxy distribution on the sky along
the line of sight. In this study, six data points of BAO dataset for $%
d_{A}(z_{\ast })/D_{V}(z_{BAO})$ are used from the Referenced. \cite{BAO1,
BAO2, BAO3, BAO4, BAO5, BAO6}. The redshift $z_{\ast }\approx 1091$ is used
as the redshift at the time of photon decoupling and $d_{A}(z)$ is the
co-moving angular diameter distance together with the dilation scale $%
D_{V}(z)=\left[ d_{A}(z)^{2}z/H(z)\right] ^{1/3}$. The chi-square function
is assumed to be \cite{BAO6}

\begin{equation}  \label{4e}
\chi _{BAO}^{2}=X^{T}C^{-1}X\,,
\end{equation}

\begin{widetext}

\begin{table}[H]
\begin{center}
\begin{tabular}{|c|c|c|c|c|c|c|}
\hline
$z_{BAO}$ & $0.106$ & $0.2$ & $0.35$ & $0.44$ & $0.6$ & $0.73$ \\ \hline
$\frac{d_{A}(z_{\ast })}{D_{V}(z_{BAO})}$ & $30.95\pm 1.46$ & $17.55\pm 0.60$
& $10.11\pm 0.37$ & $8.44\pm 0.67$ & $6.69\pm 0.33$ & $5.45\pm 0.31$ \\ 
\hline
\end{tabular}
\caption{Values of $d_{A}(z_{\ast })/D_{V}(z_{BAO})$ for different values of $z_{BAO}$.}
\label{tab1}
\end{center}
\end{table}
where 

\begin{equation*}
X=\left( 
\begin{array}{c}
\frac{d_{A}(z_{\star })}{D_{V}(0.106)}-30.95 \\ 
\frac{d_{A}(z_{\star })}{D_{V}(0.2)}-17.55 \\ 
\frac{d_{A}(z_{\star })}{D_{V}(0.35)}-10.11 \\ 
\frac{d_{A}(z_{\star })}{D_{V}(0.44)}-8.44 \\ 
\frac{d_{A}(z_{\star })}{D_{V}(0.6)}-6.69 \\ 
\frac{d_{A}(z_{\star })}{D_{V}(0.73)}-5.45%
\end{array}%
\right) \,,
\end{equation*}

and the inverse covariance matrix $C^{-1}$ is defined in
\cite{BAO6} as

\begin{equation*}
C^{-1}=\left( 
\begin{array}{cccccc}
0.48435 & -0.101383 & -0.164945 & -0.0305703 & -0.097874 & -0.106738 \\ 
-0.101383 & 3.2882 & -2.45497 & -0.0787898 & -0.252254 & -0.2751 \\ 
-0.164945 & -2.454987 & 9.55916 & -0.128187 & -0.410404 & -0.447574 \\ 
-0.0305703 & -0.0787898 & -0.128187 & 2.78728 & -2.75632 & 1.16437 \\ 
-0.097874 & -0.252254 & -0.410404 & -2.75632 & 14.9245 & -7.32441 \\ 
-0.106738 & -0.2751 & -0.447574 & 1.16437 & -7.32441 & 14.5022%
\end{array}%
\right) \,.
\end{equation*}

\begin{figure}[H]
\centering
\includegraphics[scale=0.80]{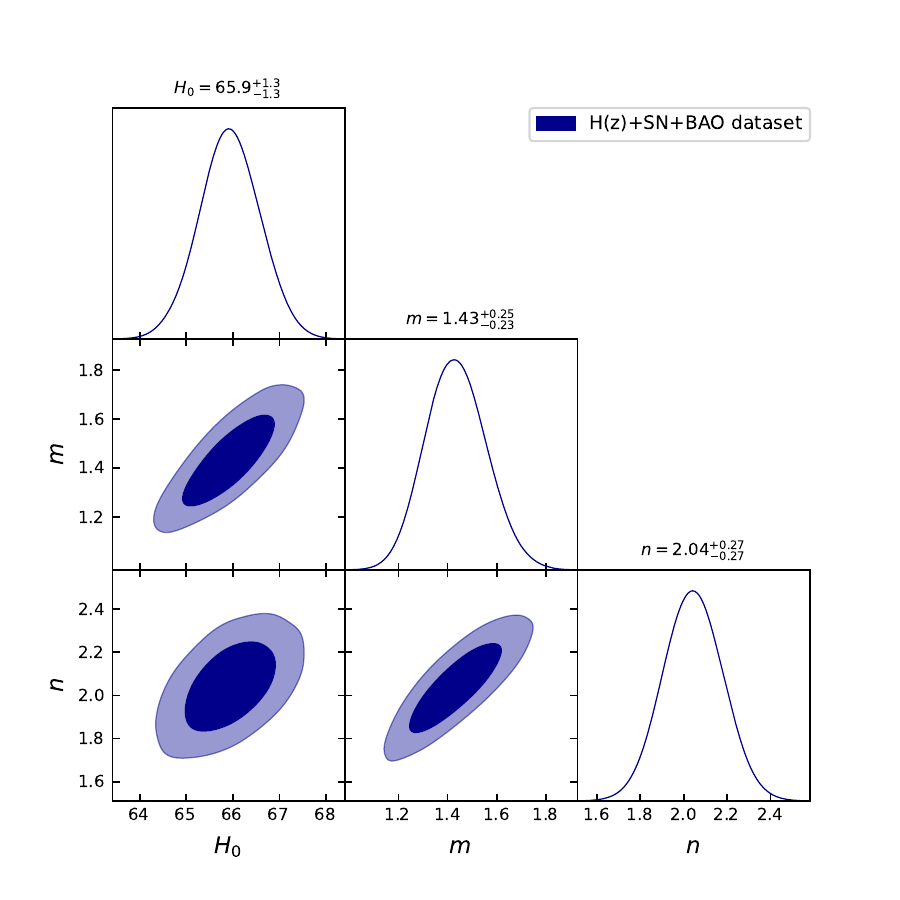}
\caption{The $1-\sigma$ and $2-\sigma$ likelihood contours for the assumed $f(Q)$ model using the $H(z)+SN+BAO$ dataset.}
\label{Cont}
\end{figure}	

\end{widetext}

For the $H(z)+SN+BAO$ dataset, the $\chi ^{2}$ function is provided as,%
\begin{equation}
\chi _{total}^{2}=\chi _{H}^{2}+\chi _{SNe}^{2}+\chi _{BAO}^{2}.
\end{equation}

By minimizing the chi-square function for the combined $H(z)+SN+BAO$ dataset, we have successfully determined the optimal values for the parameters $m$ and $n$ in our model, which provide the best fit to the data. The corresponding likelihood contours for these parameters are presented in Fig. \ref{Cont}. In this figure, the $1-\sigma$ and $2-\sigma$ likelihood contours are displayed, representing regions of parameter space that are consistent with the observed data within one and two standard deviations, respectively. These contours allow us to visualize the uncertainties associated with the model parameters and provide valuable information about the confidence intervals. Based on the likelihood contours in Fig. \ref{Cont}, we have identified the model parameters that yield the best fit to the data. The optimal values are as follows: $H_{0}=65.9_{-1.3}^{+1.3}$, $%
m=1.43_{-0.23}^{+0.25}$ and $n=2.04_{-0.27}^{+0.27}$, values indicate the central estimates for each parameter along with their associated uncertainties. The parameter $H_0$ represents the Hubble constant corresponds to the best-fit estimation obtained from the analysis of our model, as derived from the combined $H(z)+SN+BAO$ dataset. However, it is worth mentioning that this value appears to deviate from the recent Planck measurements \cite{Planck}. The values $m$ and $n$ are specific parameters within our model that have been determined to achieve the best agreement with the $H(z)+SN+BAO$ dataset. Moreover, to evaluate the statistical performance of the cosmological models, we employ well-established information criteria known as AIC (Akaike Information Criterion) \cite{Akaike:1974} and BIC (Bayesian Information Criterion) \cite{Schwarz:1974}. These criteria allow us to assess the models based on their fit to the data. AIC and BIC are calculated as follows:
\begin{eqnarray}
{\rm AIC} &=& -2 \ln {\cal L}_{\rm max}+2k\;,\nonumber\\
{\rm BIC} &=& -2 \ln {\cal L}_{\rm max}+k\ln N\;,
\end{eqnarray}
where ${\cal L}_{\rm max}$ represents the maximum likelihood, $k$ denotes the number of free parameters, and $N$ represents the total number of observational data points. The outcomes of our statistical analysis are $\chi_{min}^2=33.55$, $AIC=39.55$, $AIC_{\Lambda}=42.55$, and $BIC=45.68$. While the investigated model demonstrates lower AIC values compared to those of the standard $\Lambda$CDM model, our findings reveal that the difference in AIC values is $\Delta AIC = AIC-AIC_{\Lambda}<4$. Consequently, this indicates that the model in this study is in agreement with the expansion data, thus providing further support for their consistency.

\section{Evolution of Cosmological parameters}

\label{sec6}

The deceleration parameter is crucial in explaining the dynamics of the
Universe's expansion phase. It is determined as

\begin{equation}
q\left( z\right) =-1-\frac{\overset{.}{H}}{H^{2}}.  \label{qq}
\end{equation}

If $q>0$, our Universe is in a decelerated stage, else $q<0$ conforms to an
accelerated stage.

By using (\ref{Hz}) in (\ref{qq}), we have

\begin{equation}
q\left( z\right) =-1+\frac{3\left( 1+z\right) ^{n}}{2\left( m+\left(
1+z\right) ^{n}\right) }.  \label{qz}
\end{equation}

\begin{figure}[tbp]
{\includegraphics[scale=0.7]{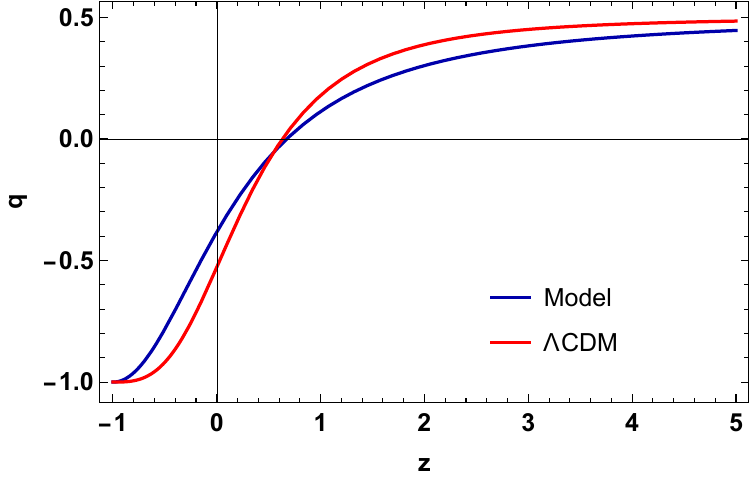}}
\caption{Evolution of the deceleration parameter for the specified model in
line with the parameter values imposed by the $H(z)+SN+BAO$ dataset.}
\label{q}
\end{figure}

From Eq. (\ref{qz}) for the deceleration parameter $q$, we see that it
contains two parameters of the model, $m$ and $n$. Fig. (\ref{q}) shows the
variation of the deceleration parameter versus redshift, and illustrates the
evolution of the Universe from the past to the present and then the future.
Taking the constrained values of the model parameters from the combined dataset used in this paper, we can see that $q$ has a transition from positive
values in the past, i.e. early deceleration, to negative values in the
present and future at the transition redshift $z_{tr}$, which leads to the current acceleration. Further, the current value of $q$ gained from the $H(z)+SN+BAO$ dataset is $q_{0}=-0.38_{-0.06}^{+0.06}$ , which is consistent with the observational data. The value of the transition redshif is $z_{tr}=0.68_{-0.01}^{+0.01}$.

\begin{figure}[tbp]
{\includegraphics[scale=0.7]{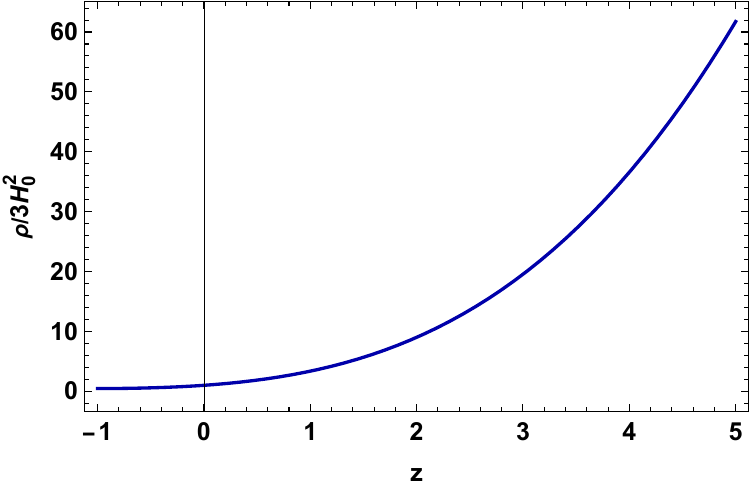}}
\caption{Evolution of the energy density for the specified model in line
with the parameter values imposed by the combined $H(z)+SN+BAO$ dataset.}
\label{rho}
\end{figure}

Figs. \ref{rho} shows the behavior of the energy density of the Universe. It
has been shown that the energy density of the Universe increases positively with redshift.

\begin{figure}[tbp]
{\includegraphics[scale=0.7]{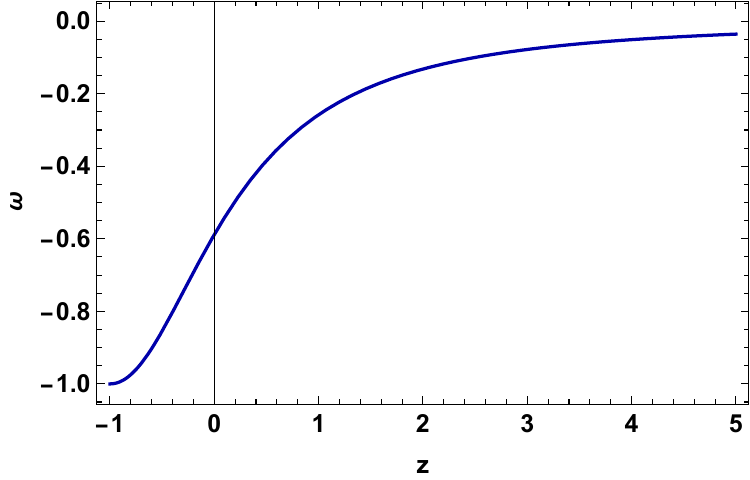}}
\caption{Evolution of the EoS parameter for the specified model in line with
the parameter values imposed by the combined $H(z)+SN+BAO$ dataset.}
\label{omega}
\end{figure}

A good method to understand the existence of DE and the accelerating phase
of the Universe is to determine the value and evolution of the Equation of
State (EoS) parameter. The EoS parameter is defined as $\omega =\frac{p}{%
\rho }$, where $p$ and $\rho $ represent the isotropic pressure and energy
density of the Universe, respectively. According to recent Planck data
(2018) \cite{Planck}, the present value of EoS is $\omega _{0}=-1.028\pm
0.032$. It can be argued that the observational evidence supports the
phantom model of dark energy ($\omega <-1$). Also, there are other
candidates on dark energy such as the cosmological constant ($\omega =-1$)
and quintessence dark energy ($-1<\omega <-\frac{1}{3}$). The EoS parameter
can provide information on the history of the expansion of the Universe over
cosmic time, if $\omega =\frac{1}{3}$ the Universe is dominated by
radiation, and if $\omega =0$ is called the matter-dominated era, while in
the case of $\omega <0$ it is called the dark energy-dominated Universe.

From Eq. (\ref{EoS}), we can see that the EoS parameter of our cosmological $%
f\left( Q\right) $ model is time-dependent and converges to $\omega
\rightarrow -1$ as cosmic time increases. In line with constrained values of
model parameters from the $H+SN+BAO$ dataset, the variation of the EoS
parameter versus redshift $z$ is exhibited in Fig. \ref{omega}. It is noted
that the EoS parameter of our cosmological $f\left( Q\right) $ model varies
in the quintessence region for the constrained value of the model parameters
from the combined dataset. Further, it is clear that the $\omega $
approaches $\Lambda $CDM model in the future. The present value of EoS
parameter is obtained as $\omega _{0}=-0.59_{-0.04}^{+0.04}$ for the $H+SN+BAO$ dataset, which approximates the measured value from Planck 2018 results \cite{Planck}.

\section{Energy conditions}

\label{sec7}

The energy conditions (ECs), which are based on the Raychaudhuri equation,
are crucial for describing the behavior of the compatibility of timelike,
lightlike, or spacelike curves and are frequently utilized to comprehend
terrifying singularities in black holes \cite{ECs}. Koussour et al. \cite{A27}
has been used to study the ECs in symmetric teleparallel gravity. ECs also
help us check the validity of the model. In particular, the violation of a
strong energy condition leads to an acceleration of the Universe. The ECs
are described as:

\begin{itemize}
\item Weak energy condition (WEC): $\rho +p\geq 0$ and $\rho \geq 0$;

\item Null energy condition (NEC): $\rho +p\geq 0$;

\item Dominant energy condition (DEC): $\rho \geq \left\vert p\right\vert $
and $\rho \geq 0$;

\item Strong energy condition (SEC): $\rho +3p-6\overset{.}{f}_{Q}H+f\geq 0$%
. 
\begin{figure}[tbp]
{\includegraphics[scale=0.70]{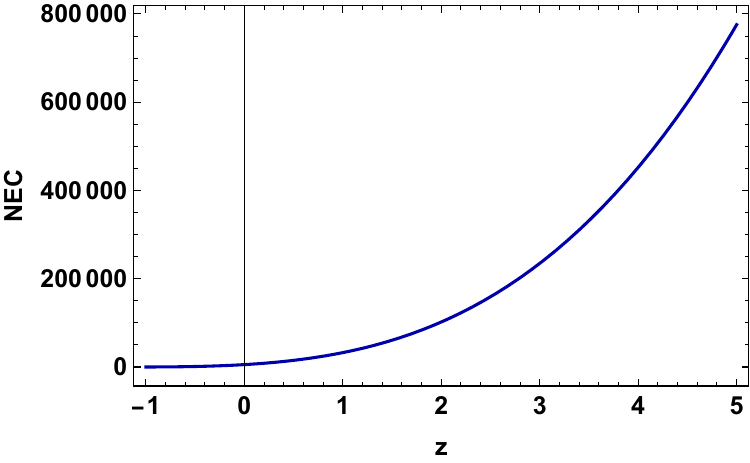}}
\caption{Evolution of the NEC for the specified model in line with the
parameter values imposed by the combined $H(z)+SN+BAO$ dataset.}
\label{NEC}
\end{figure}

\begin{figure}[tbp]
{\includegraphics[scale=0.70]{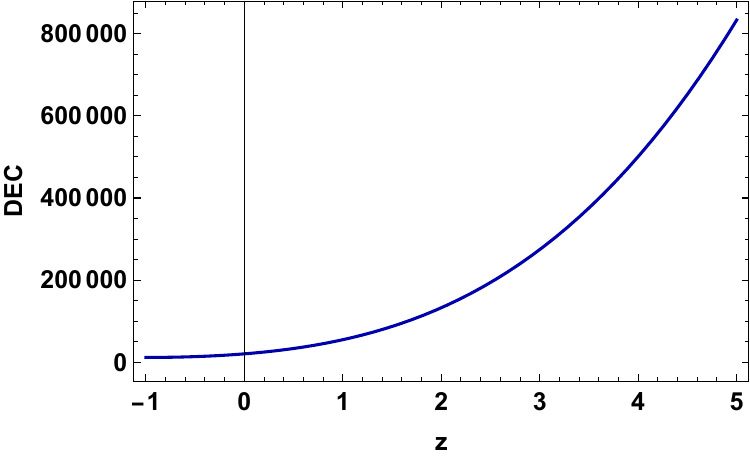}}
\caption{Evolution of the DEC for the specified model in line with the
parameter values imposed by the combined $H(z)+SN+BAO$ dataset.}
\label{DEC}
\end{figure}

\begin{figure}[tbp]
{\includegraphics[scale=0.70]{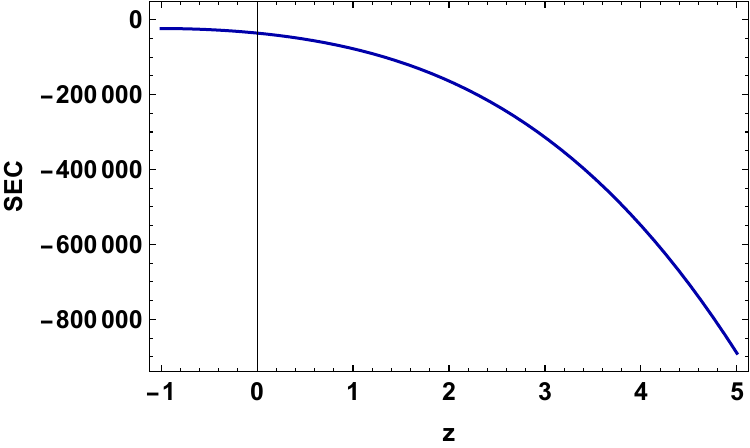}}
\caption{Evolution of the SEC for the specified model in line with the
parameter values imposed by the combined $H(z)+SN+BAO$ dataset.}
\label{SEC}
\end{figure}
\end{itemize}

It is evident from Fig. \ref{rho} that the WEC condition i.e. energy density
displays favorable behavior. Further, we observed from Figs. \ref{NEC}, \ref%
{DEC} and \ref{SEC} that the NEC and DEC conditions are met, but the SEC
condition is not. As we noted previously, violating the SEC condition causes
the Universe to accelerate. The conclusions from our cosmological model
agree with several works in the literature \cite{A22,A23}.

\section{Diagnostic analysis}

\label{sec8}

\subsection{Statefinder diagnostic}

For analyzing or characterizing cosmic acceleration, more and more DE models are being built. A sensitive and reliable diagnostic for DE models is
necessary to be able to distinguish between these conflicting cosmological
scenarios employing DE. For this reason, Sahni et al. developed a diagnostic
approach that utilizes the parameter pair $\left( r,s\right) $, the
so-called "statefinder" \cite{Sahni}. It is simple to see that the
statefinder is a logical progression from $H$ and $q$ and probes the
dynamics of the Universe's expansion through higher derivatives of the scale
factor. The statefinder diagnostic is a helpful tool for differentiating
these DE models since different cosmological models including DE exhibit
qualitatively distinct evolution paths in the $r-s$ plane. The statefinder
parameters for the spatially flat $\Lambda $CDM model are ($r=1,s=0$). In
addition, Quintessence model relates to ($r<1,s>0$), Chaplygin gas model to (%
$r>1,s<0$), and Holographic DE model to ($r=1,s=\frac{2}{3}$). The following
definition applies to the statefinder pair $\left( r,s\right) $:%
\begin{equation}
r=\frac{\overset{...}{a}}{aH^{3}},
\end{equation}%
\begin{equation}
s=\frac{\left( r-1\right) }{3\left( q-\frac{1}{2}\right) }.
\end{equation}

The parameter $r$ can be rewritten as%
\begin{equation*}
r=2q^{2}+q-\frac{\overset{.}{q}}{H}.
\end{equation*}%
\begin{figure}[tbp]
{\includegraphics[scale=0.7]{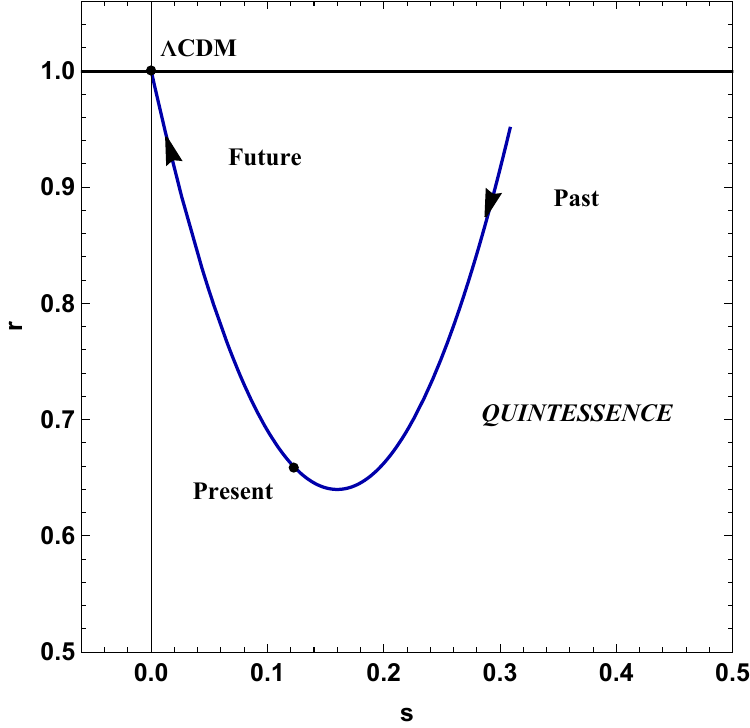}}
\caption{Evolution of the $r-s$ plane for the specified model in line with
the parameter values imposed by the combined $H(z)+SN+BAO$ dataset.}
\label{rs}
\end{figure}

Fig. \ref{rs} represents the $r-s$ plane by considering parameters
constrained using the H(z)+SN+BAO dataset. The provided model currently
belongs to the quintessence zone. We can see that in the
future, the trajectories of the $r-s$ graph will pass via the $\Lambda $CDM
fixed point.

\subsection{$Om$ diagnostic}

\begin{figure}[tbp]
{\includegraphics[scale=0.7]{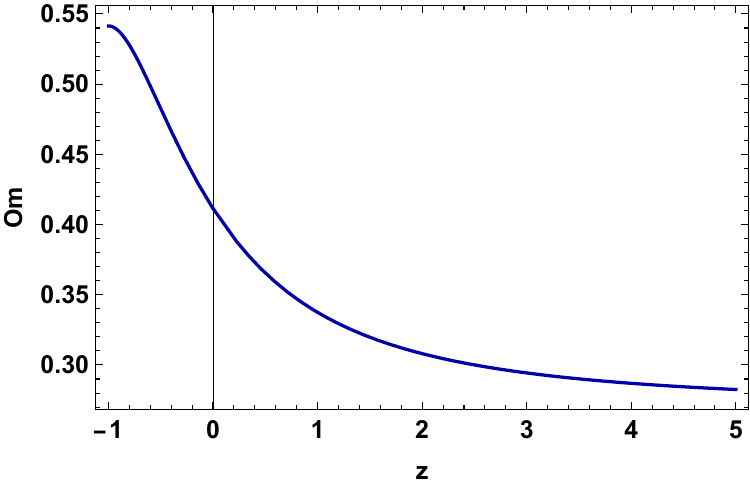}}
\caption{Evolution of the $Om(z)$ diagnostic for the specified model in line
with the parameter values imposed by the combined $H+SN+BAO$ dataset.}
\label{om}
\end{figure}

In this subsection, we explore the Om diagnostic, another crucial tool for
examining the dynamical nature of cosmological models that deal with the
problem of dark energy \cite{Sahni1}. Since it only employs the Hubble
parameter $H$, it is a simpler diagnostic than the statefinder diagnostic
explained in the previous section. The Om diagnostic is defined as the
following in a spatially flat Universe: 
\begin{equation}
Om\left( z\right) =\frac{\left( \frac{H\left( z\right) }{H_{0}}\right) ^{2}-1%
}{\left( 1+z\right) ^{3}-1}
\end{equation}%
where $H_{0}$ is the current value of the Hubble constant. The model
exhibits quintessence behavior with a negative slope of $Om(z)$, whereas a
positive slope denotes the model's phantom behavior. Lastly, the constant $%
Om(z)$ behavior alludes to the $\Lambda $CDM model. From Fig. \ref{om}, it
is clear that $Om(z)$ has a negative slope which indicates that our
cosmological model represents quintessence-type behavior. 

\section{Conclusion}

\label{sec9}

The problem of late-time cosmic acceleration could not be adequately
addressed in the context of GR, but geometrically modified gravity might be
able to shed some light on this contemporary cosmology problem. According to
GR, this acceleration results from the dark energy, or strongly negative
pressure, fulfilling $\rho \left( 1+3\omega \right) <0$. This article
examined a cosmological model in which the phenomenon of cosmic acceleration
was described without the necessity for dark energy. By using the
parametrization method, the dynamical equation of the Hubble parameter is
presumed i.e. Eq. (\ref{Hz}) in modified $f(Q)$ gravity with the FLRW
background. The model we looked at, $f\left( Q\right) =- Q+\frac{\alpha 
}{Q}$, comprises both a linear and a non-linear type of non-metricity, where  $\alpha $ is the free model parameter. In addition, we used the
combined $H(z)+SN+BAO$ dataset to find the best-fit values of the model
parameters. The results of the best fit are $H_{0}=65.9_{-1.3}^{+1.3}$, $%
m=1.43_{-0.23}^{+0.25}$ and $n=2.04_{-0.27}^{+0.27}$. In addition, we have
discussed the behavior of the various cosmological parameters obtained in
order to analyze the detailed evolution of the Universe in various periods.
The evolution of the deceleration parameter in Fig. \ref{q} shows that the
Universe recently switched from a decelerated to an accelerated stage, while
the energy density in Fig. \ref{rho} behaves as expected. The current value
of the deceleration parameter is $q_{0}=-0.38_{-0.06}^{+0.06}$ for the
$H(z)+SN+BAO$ dataset. Also, the value of the transition redshif is $z_{tr}=0.68_{-0.01}^{+0.01}$. In Fig. \ref{omega}, the evolution of the equation of
state parameter is depicted. The negative value of the $\omega _{0}=-0.59_{-0.04}^{+0.04}$ for the $H+SN+BAO$ dataset is seen to remain in the quintessence
zone and approach to $-1$ in the far future, resulting to the $\Lambda $%
CDM model (i.e., it does not pass the phantom division line at $\omega <-1$%
). The evolution of the energy conditions has been covered. It is clear that
while NEC and DEC (see Figs. \ref{NEC} and \ref{DEC}) do satisfy the model
but SEC (see Fig. \ref{SEC}) is not satisfied, creating an attractive force
that places the Universe in an accelerating phase. Finally, an examination
of the statefinder parameters and the $Om$ diagnosis has been performed as
well as compared to the $\Lambda $CDM model. 

The model described here is a
straightforward explanation of how cosmic evolution behaves in the late Universe under modified gravity. It shows quintessence-like behavior in the present and shows a nice fit to some observational datasets, making our model a suitable alternative to standard lore. \newline

\section*{Acknowledgments}
This research is funded by the Science Committee of the Ministry of Science and Higher Education of the Republic of Kazakhstan (Grant No. AP09058240).

\textbf{Data availability} There are no new data associated with this
article.


\end{document}